\documentclass[times, twoside]{zHenriquesLab-StyleBioRxiv}
\usepackage{blindtext}
\usepackage{graphicx}

\leadauthor{Preoteasa} 

\begin{document}

\title{EMERALD-UI: An interactive web application to unveil novel protein biology hidden in the suboptimal-alignment space}
\shorttitle{Visualising Suboptimal Protein Alignments}

\author[a]{Andrei Preoteasa}
\author[a,b]{Andreas Grigorjew}
\author[a]{Alexandru I. Tomescu*}
\author[c,d]{Hajk-Georg Drost*}

\affil[a]{Department of Computer Science, University of Helsinki, 00560 Helsinki, Finland}
\affil[b]{LAMSADE, CNRS UMR7243, Université Paris Dauphine-PSL, 75775 Paris, France}
\affil[c]{Computational Biology Group, Max Planck Institute for Biology Tuebingen, Germany}
\affil[d]{Digital Biology Group, Faculty of Life Sciences, University of Dundee, UK}

\maketitle

%TC:break Abstract
%the command above serves to have a word count for the abstract
\begin{abstract}

Life over the past four billion years has been shaped by proteins and their capacity to assemble into three dimensional conformations. Protein sequence alignments have been the enabling technology for exploring the evolution and functional adaptation of proteins across the tree of life. Recent advancements in scaling the prediction of three dimensional protein structures from primary sequence alone, revealed that different modes of conservation and function operate on the sequence and structure level. This difference in protein conservation patterns and their underlying functional change that could emerge in suboptimal alignment configurations is often ignored in optimal protein alignment approaches. We introduce EMERALD-UI, an open-source interactive web application which is designed to reveal unexplored biology by visualising stable structural conformations or protein regions hidden in the suboptimal alignment space.\\

\textbf{Availability}: EMERALD-UI is available at \url{https://algbio.github.io/emerald-ui/}.\\

\textbf{Contact}: \href{email:hdrost001@dundee.ac.uk}{hdrost001@dundee.ac.uk} or \href{email:alexandru.tomescu@helsinki.fi}{alexandru.tomescu@helsinki.fi}. 

\end {abstract}
%TC:break main
%the command above serves to have a word count for the abstract


\begin{corrauthor}

\end{corrauthor}

\section*{Introduction}
When aligning two protein sequences, a user defined (and often heuristic) cost-matrix determines which alignment configuration is deemed most optimal by a respective alignment algorithm \citep{zahid2015novel,petti2023end,llinares2022deep,kaminski2022plm}. For two divergent protein sequences, for example, this could mean that thousands of alternative alignment configurations exist, but with lower total cost-matrix score. The core assumption of this optimal alignment approach is that due to its well-established heuristic insight, the optimal score is proportional to a biologically meaningful alignment configuration, especially when dealing with similar proteins with greater than 90\% sequence identity \citep{naor1994near}. Although this assumption seems reasonable, when extrapolated to millions of pairwise comparisons, this approach can introduce a systematic optimal alignment bias, where alternative alignment configurations would have been more reasonable when biologically assessed \citep{emerald}. We recently introduced the open-source command line tool EMERALD which allows users to systematically explore the suboptimal alignment space and sample alignment-safe positions from the theoretically true alignment configuration \citep{emerald}. Using EMERALD, users can extract individual alignment-configuration positions which are stable across all or a proportion of suboptimal alignments. However, this tool lacks a comprehensive user interface and analytics capacity to reveal alternative biology or protein structural information hidden in suboptimal alignment configurations.

% footnote\footnote{\url{https://algbio.github.io/emerald-ui/}}
EMERALD-UI solves this problem by providing an interactive web application to visualise the suboptimal alignment space between two protein sequences and map alignment-stable positions onto the three dimensional structure on the respective protein folds. We envision that users will be able to trial how alternative (suboptimal) alignment configurations will translate into differences in protein structural or protein-protein interaction space. Especially, when deployed to proteins involved in protein complexes, we foresee new biological and functional insights emerging from new (and previously unrecognised) configurations of protein complexes that will arise from the systematic exploration of the suboptimal alignment space. Particularly when dealing with the comparison of highly divergent proteins, the suboptimal alignment space will reveal an evolutionary repertoire of structural conformations which can also be investigated in the context of their role in biological pathways, kinetic efficiency (e.g. enzymes) or drug target predictions.

\section*{EMERALD: As Standalone Command Line Tool}

In \citep{emerald}, we introduced the notion of \emph{alignment-safety} for pairwise protein sequence alignments: EMERALD explicitly calculates the space of alternative (suboptimal) alignments. From this space, a set of \emph{safety windows} can be extracted as contiguous alignment intervals that are shared across many (or all) alternative alignments. 

In detail, EMERALD considers a standard directed acyclic graph (DAG) representing a dynamic-programming alignment (using Needleman–Wunsch alignment with affine gaps)~\citep{likic2008needleman}. To also represent sub-optimal alignments, this DAG incorporates all edges contained in an alignment of cost at most $\Delta$ from the optimal one. In this graph, EMERALD computes alignment intervals (safety windows) that are common to an $\alpha$ proportion of all  alignments in this DAG ($\alpha = 1$ means common to \emph{all} alignments). These safety windows are projected back onto the coordinates in the two protein sequences. 

In a large‑scale experiment using the SwissProt database, we showed that especially for divergent sequence pairs, safety windows can retain a high fraction of structurally stable protein residues while covering only a small fraction of the sequence \citep{emerald}. This result supports the idea that exploring suboptimal alignment space reveals structurally significant conserved regions that a single optimal alignment would otherwise miss.

In practice, the two EMERALD parameters $\Delta$ and $\alpha$ impact alignment-safety computations as follows: 
\begin{itemize}
    \item Increasing $\Delta$ enlarges the suboptimal alignment space. Moreover, as $\Delta$ increases, safety windows typically become shorter and/or fewer (and overall ``safe coverage'' tends to drop), because more alternative alignments create more disagreement about which residue pairings are robust.
    \item Increasing $\alpha$ makes the safety criterion stricter, meaning that safety-windows tend to become shorter or fewer, but result in higher-confidence. Decreasing $\alpha$ on the other hand makes the results more permissive, yielding longer or more windows, but less stable.   
\end{itemize}

\section*{EMERALD-UI: An Interactive Web-Application}
 
 The new EMERALD-UI web application provides an interactive and responsive user interface for generating and exploring the suboptimal protein alignment space when comparing two protein sequences. Moreover, while EMERALD as a standalone command line tool requires user installation, EMERALD-UI is running natively in any modern web-browser environment without manual installation requirements. 
 
\begin{figure*}[t!]%
\centering
% \includegraphics[width=\textwidth]{img/EMERALD-UI.pdf}
% \caption{test}\label{fig1}
\includegraphics[width=\textwidth]{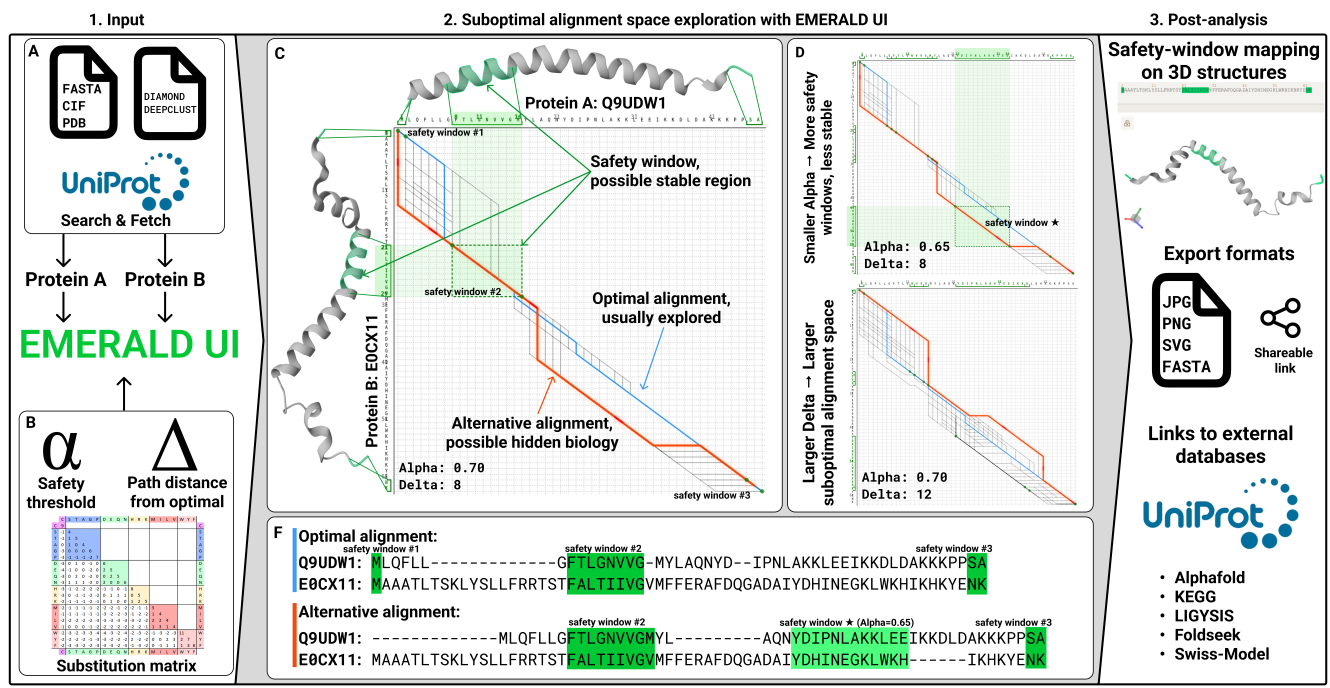}
\caption{
\textbf{Schematic Overview of the EMERALD-UI Workflow.} 1) Users can import their protein sequences as FASTA, CIF, PDB, Uniprot and DIAMOND DeepClust clusters for comparisons across the suboptimal alignment space (panel A). 
2)  Using the protein sequences specified by the user, EMERALD-UI then computes the pairwise (suboptimal) alignments using EMERALD either by relying on the default settings or the users adjusted (or even advanced) settings (panel B) and visualises the corresponding alignment graph which includes all alternative alignment configurations of the suboptimal alignment space (panel C).  When lowering the alpha parameter new safety windows across the suboptimal alignment space are revealed. This is captured by paths diverting from the optimal alignment path (blue path). When increasing the delta parameter on the other hand, more alternative alignment paths are revealed. Users can then select an alternative alignment path (red path) for further structural and downstream assessments (panel D). 3) Finally, users can explore which parts of their protein sequence and corresponding structure are alignment safe (stable across suboptimal alignment configurations) (panel F) and forward these proteins to downstream protein annotation databases for further exploration.
Alignment available at: \url{https://algbio.github.io/emerald-ui/?seqA=E0CX11&seqB=Q9UDW1}}\label{fig1}
\end{figure*}

 The UI takes two user defined protein sequences as input and internally runs EMERALD directly in the browser environment to ultimately display the suboptimal alignment results as well as their associated protein structures with safty-windows projected onto the structure model. All results generated with EMERALD-UI can be exported to the user machine. EMERALD-UI also provides direct URL links to downstream databases such as UniProt~\citep{Uniprot2025}, AlphaFold~\citep{alphafold2021, alphafold2023}, FoldSeek~\citep{Foldseek2024}, KEGG~\citep{Kegg2000,Kegg2019,Kegg2024}, LigySis~\citep{Ligysis2024,Ligysis2024*,Ligysis2025} for further inspection of how suboptimal alignment configurations translate into differences in structural space (Fig.~\ref{fig1}).
 
 The application supports multiple input methods. The user can either upload their own protein sequence files in fasta-format, manually input protein sequences by copy pasting, upload PDB or CDF files, or search the Uniprot database directly from EMERALD-UI.
 
 With the protein pair defined, the user can then fine tune EMERALD parameters. The $\alpha$ slider changes the safety window highlighting threshold. The $\delta$ slider changes the suboptimal alignment depth. The input also includes advanced parameters which can be used to change the EMERALD alignment cost matrix, gap cost and gap-opening cost. We support BLOSSUM 45, 50, 62, 80, 90, PAM 30, 70, 250 and basic IDENTITY. With these advanced parameters the user can explore how the suboptimal alignment space and corresponding structural changes change when different evolutionary scenarios (different cost-matrices) are employed.

For each case, the generated alignment is visualised in a custom graph drawing. An alignment of characters is visualised by a diagonal line between them, whilst a mismatch is seen as either a vertical or horizontal line. The alignment safety windows are highlighted in green within the graph. To illustrate alternative suboptimal alignment configurations, the graph highlights these alternative configurations as different alignment paths along the graph. The ``traditional'' optimal alignment is visualised in blue. The graph is highly interactive so that users can zoom in and out of the graph to select their preferred path and alternative alignment configuration. Hovering over any section of the alignment graph shows the characters at that position, their positional indices and fraction of how many (sub)optimal alignment paths go through that position. Further analysis can be done through the side panel which lets the user highlight and view information about safety windows and non-safety-windows. The side panel is also designed to encourage users to select a specific sub-path and thereby explore specific subsections of the suboptimal alignment space. The graph can be exported directly as a PNG, JPEG or SVG file and used in publications. The side panel includes a settings tab where users can toggle different features in the graph on and off depending on which features they wish to visualise in the exported image.

Finally, EMERALD-UI connects any suboptimal alignment assessment to downstream databases for further exploration of how the suboptimal alignmentspace can fuel novel discovery in the structural and biological pathway space. For example, when using proteins from UniProt, EMERALD-UI projects the generated safety windows directly onto the structure of the protein through Mol* (industry standard for 3D visualization) \cite{mol*2021} and links the respective proteins to KEGG and LIGSYS for downstream pathway and ligandability assessments. 

\section*{Code Highlights and Software Architecture}
EMERALD-UI has a simple and lightweight structure, which enables real-time and no-delay execution without local installation or server deployment. The application uses EMERALD natively in the browser through Web Assembly (WASM). The raw output of EMERALD is compiled in real time and then displayed as a graph. The graph rendering and deployment logic was written ground-up with the D3.js JavaScript library, meaning that exploring the graph (zooming and panning) is fast and smooth even with large inputs. The graph is interactive and lets the user explore and highlight specific sections of the alignment. Each safety window can be highlighted and viewed separately in a side panel (see documentation on EMERALD-UI webpage). In addition to the optimal alignment being highlighted in the graph (in blue) it is also shown as a traditional pairwise alignment configuration below the alignment graph. If the user selects a custom path in the graph (in orange), the corresponding alternative alignment is also shown as a pairwise alignment in the same panel as the optimal pairwise alignment. 

EMERALD-UI also supports exporting different data in different formats. If the user has selected sections in the graph, i.e. a safety window, a non-safety window or a full alignment, they can be exported as raw text directly from the application. The graph itself can be exported as an image file. The alignment can also be directly shared as a URL, which when opened, takes the user directly to the exact same protein alignment.

The protein retrieval tool embedded in EMERALD-UI pulls the latest protein data through the UniProt API. The search box supports searching proteins by scientific names, accession codes or ID codes and even natural language. The search results are compiled into a list table with the accession code, protein name, organism and length. The user is also given the option to preview the protein, navigate to the UniProt page of the protein and further explore the corresponding 3D structure of the protein through Mol*.

\section*{Code Availability}
EMERALD-UI is available at \url{https://algbio.github.io/emerald-ui/}, and its underlying source code can be found at \url{https://github.com/algbio/emerald-ui}. The original EMERALD Command Line Tool~\citep{emerald} can be found at \url{https://github.com/algbio/emerald}.

\section*{Acknowledgments}

We thank Sebastian Schmidt for discussions on EMERALD-UI. We also thank all members of the DrostLab and Algorithmic Bioinformatics Lab for their feedback on user-experience and overall UI features. This work was supported by the Max Planck Society and a Royal Society Wolfson Fellowship (RSWF\textbackslash R1\textbackslash 241004 awarded to H.G.D). H.G.D also thanks Detlef Weigel for generous sponsorship and the BMBF-funded de.NBI Cloud within the German Network for Bioinformatics Infrastructure (de.NBI) (031A532B, 031A533A, 031A533B, 031A534A, 031A535A, 031A537A, 031A537B, 031A537C, 031A537D, 031A538A) for computational support. This work was partially funded by the Research Council of Finland (grants No.~346968, 358744) and by the European Union (ERC, SCALEBIO, 101169716, awarded to A.I.T). Views and opinions expressed are however those of the author(s) only and do not necessarily reflect those of the European Union or the European Research Council. Neither the European Union nor the granting authority can be held responsible for them. Finally, A.G. is partially supported by the ANR project ANR-21-CE48-0022 (S-EX-AP-PE-AL), awarded to Michael Lampis.

\section*{Bibliography}
\bibliography{reference}

%% You can use these special %TC: tags to ignore certain parts of the text.
%TC:ignore
%the command above ignores this section for word count
\onecolumn

\end{document}